\UseRawInputEncoding
\documentclass[aps,prb,twocolumn,superscriptaddress]{revtex4-1}
\usepackage{graphicx,bm,amssymb,amsmath,xcolor,pifont,soul}
\usepackage[linktocpage=true,colorlinks=true,pdfborder={0 0 0},linkcolor=blue,citecolor=red,filecolor=yellow,urlcolor=blue,bookmarks,pdfauthor={},]{hyperref}
\usepackage{ulem}
\usepackage{multirow}

\begin{document}

\title{\texorpdfstring{Singly occupied 4$f$ antiferromagnetic insulators: CePO$_4$ and CeVO$_4$}
                     {Singly occupied 4f antiferromagnetic insulators: CePO4 and CeVO4}}

\author{Hari Paudyal}
\affiliation{Department of Physics and Astronomy, University of Iowa, Iowa City, Iowa 52242, USA}
\author{Yogendra Limbu}
\affiliation{Department of Physics and Astronomy, University of Iowa, Iowa City, Iowa 52242, USA}
\author{Michael E. Flatt\'e}
\affiliation{Department of Physics and Astronomy, University of Iowa, Iowa City, Iowa 52242, USA}
\affiliation{Department of Applied Physics, Eindhoven University of Technology, Eindhoven, The Netherlands}
\author{Durga Paudyal}
\affiliation{Department of Physics and Astronomy, University of Iowa, Iowa City, Iowa 52242, USA}
\date{\today}
\begin{abstract}
Rare-earth containing wide band gap oxides, which provide spin-photon interface and narrow linewidth optical emission, are getting significant attention as the most promising candidate materials in advancing quantum transduction and memories. Here, from $ab~initio$ calculations, we identify antiferromagnetic ground states in structurally preferred monoclinic CePO$_4$ and tetragonal CeVO$_4$ exhibiting localized occupied and unoccupied Ce $4f$ states with $4f-4f$ transition characteristics. Interestingly, in CePO$_4$, O $2p$ and P $3p$ states hybridize negligibly with Ce $4f$ states, while in CeVO$_4$, V $3d$ and O $2p$ states hybridize and appear as extended states in between the occupied and unoccupied Ce $4f$ states. 
Here, phonon calculations and analysis identify and differentiate Raman active phonon modes along with the spin phonon coupling of Ce in both CePO$_4$ and CeVO$_4$ that ultimately lead to different $4f$ ground state crystal field multiplets, which are critical to accurately describe electronic transitions for foundational quantum transduction and memories. Further, the identified $C_1$ site symmetry of Ce, lacking inversion symmetry in CePO$_4$, is relevant for quantum memories and $D_{2d}$ site symmetry of Ce exhibiting inversion symmetry in CeVO$_4$ is relevant for quantum transduction.

\end{abstract}
\maketitle
\section{Introduction}
Rare-earth (RE) containing materials exhibit a wide range of fascinating electronic and magnetic ground and excited state properties due to the interplay between electronic spins and crystal lattices ~\cite{heshami2016quantum, kolesov2012optical, siyushev2014coherent}. The partially filled 4$f$ and 5$d$ states in these materials contribute to possess complex magnetic states, strong spin-orbit coupling (SOC), and sharp electronic excitations that drive to unique ground and excited states relevant for quantum information science (QIS)~\cite{kolesov2013mapping, savchenkov2023uncommon}. RE spins are notably intriguing because they can display precise optical and microwave transitions and possess substantial magnetic dipole moments~\cite{everts2020ultrastrong}. A particular attention has been given to stoichiometric RE oxides, such as ortho-phosphate, ortho-vanadate, ortho-silicate, and ortho-tungstate in interfacing spins and photons for quantum information processing~\cite{xie2021characterization, rakonjac2020long}. This is mainly due to the location of the well shielded RE 4$f$ states in these wide bandgap oxides providing distinct and precisely defined energy levels with relatively low interaction between the electronic and vibrational wave functions. Within this group of RE oxides, cerium based ones stand out due to the presence of a single 4$f$ electron. For instance, cerium ortho-phosphate (CePO$_4$) and cerium ortho-vanadate (CeVO$_4$) have Ce in 3+ valance state with only one optically active 4$f$ electron exhibiting both emission and absorption spectra~\cite{xie2021characterization, rakonjac2020long}.

In the past, REPO$_4$ and REVO$_4$ were used in laser optics and electro- and photo-catalysis applications. These materials, as renaissance quantum materials, may have applications as optical transmission medium in quantum photonics. Specifically, CePO$_4$ nanotubes and CeVO$_4$ heterojunctions for quantum dots have been employed in photoemissive devices~\cite{tang05nanotube}, leading towards photonic communication in QIS. Interestingly, both CePO$_4$ and CeVO$_4$ exhibit antiferromagnetic (AFM) ground states: former possessing wider band-gap by $\sim$3~eV as compared to the latter, showing a promise as an excellent medium for  solid-state spin qubits~\cite{wolfowicz2021quantum}.

It is crucial that high-quality wide-bandgap crystals at cryogenic temperatures should be free from decoherence. One of the decohering factors is nearby nuclear spins, suggesting the maximization of spin coherence in an optimal host material with near-zero nuclear spins. Another important decohering factor depends on the magnetic configurations governed mainly by states hybridization. As for example, lacking of $3d$ states in CePO$_4$ may provide preference of Ce spins align antiferromagnetically, whereas hybridization between V $3d$ and Ce $4f$/$5d$ states may induce magnetic noise in CeVO$_4$ due to hybridized states in between occupied and unoccupied 4$f$ density of states (DOS) peaks. It is apparent that CePO$_4$ can be a better material for quantum information processing with a possibility of no intrinsic magnetic noise, with a caveat of low symmetry ground state configuration.  Indeed, AFM RE systems hold an ability to generate no stray magnetic field making them resilient to external magnetic perturbations and exhibiting a narrow, shape independent resonant linewidths~\cite{everts2020ultrastrong}. These characteristics make AFM systems robust to external magnetic perturbations ensuring greater stability in magnetic environments. Ferromagnetic (FM) systems, on the other hand, lead to an interference due to stray magnetic field and are limited for certain quantum applications~\cite{jenkins2019magnetic}.

The Ce $4f$ and $5d$ states in CePO$_4$ and CeVO$_4$ form ground and excited state multiplets leading to absorption and emission spectra depending on their crystal symmetry. These spectra are observed from Raman spectroscopy measurements and also calculated from phonon dispersions. The important aspect of this is the vibrational transition polarizability, which is significantly amplified when the excitation resonates with the ground and excited state transitions. Therefore, it is crucial to understand phonon dynamics and its possible connections to the quantum transduction and quantum memories.

Since, Ce $4f$ states are correlated electronic states, a special treatment is needed. The standard density functional theory (DFT) approaches fail owing to significant self-interaction errors associated with $4f$ states~\cite{adelstein2011structure, perdew1981self-interaction}. This failure even introduces problems in predicting states of matter mostly in semiconducting or insulating materials~\cite{alvarez2023strongly}. A commonly used alternative to partially fix such error is by adding onsite electron correlation, Hubbard $U$, in DFT~\cite{dudarev1998electron, anisimov1991band}, however, calibration of the $U$ value and its validation inevitably reduces the predictive power of the method. Hybrid functionals, introducing a fraction of exact exchange, which go beyond such parameter dependent Hubbard model, improve the description of $d$- and $f$-electron behavior.

In this paper, we first perform $ab~initio$ calculations to identify AFM ground states in structurally preferred monoclinic CePO$_4$ and tetragonal CeVO$_4$, which exhibit localized occupied and unoccupied Ce $4f$ states with Ce $4f$ - Ce $4f$ transition characteristics. O $2p$ and P $3p$ states hybridize negligibly with Ce $4f$ in CePO$_4$ contrasting with the hybridized V $3d$ and O $2p$ states as extended states in between the occupied and unoccupied Ce $4f$ states in CeVO$_4$. Next, we perform phonon calculations to identify and differentiate Raman active phonon modes along with spin phonon coupling of Ce in both CePO$_4$ and CeVO$_4$ that ultimately lead to different $4f$ ground state crystal field multiplets. The $C_1$ site symmetry of Ce in CePO$_4$ lacks inversion symmetry and $D_{2d}$ site symmetry of Ce in CeVO$_4$ does include inversion symmetry, and therefore, CePO$_4$ can be used for quantum memory and CeVO$_4$ can be used for quantum transduction.

\section{Computational methods}
\textit{Ab initio} calculations are performed using Vienna \textit{Ab initio} Simulation Package~\cite{furthmuller1996dimer, kresse1993ab, kresse1996efficiency} with projector augmented wave potentials~\cite{blochl1994projector} within the Perdew-Burke-Ernzerhof~\cite{perdew1996generalized} exchange-correlation functionals in the generalized gradient approximation (GGA). An optimized plane wave kinetic-energy  cutoff value of 500~eV, a $\Gamma$-centered $6 \times 6 \times 6$ Monkhorst-Pack \textbf{k}-mesh~\cite{Monkhors1976s}, and a Gaussian smearing width of 0.01~eV are used for the Brillouin-zone sampling. The atomic positions and lattice parameters are optimized until the self-consistent energy is reached within 10$^{-5}$~eV and the maximum force on each atom is less than 10$^{-4}$~eV/\AA. For the DOS, a denser \textbf{k}-mesh of $12 \times 12 \times 12$ is used. The SOC energy and orbital magnetic moments are calculated using noncollinear spin-orbit calculations. To understand the correlated electron magnetism, spin polarized electronic structure calculations are performed incorporating various electron-electron correlation parameters, $U_{eff}=U-J$ = 4 to 6~eV, where $U$ and $J$ are the onsite Coulomb repulsion and exchange interaction. These values are consistent with the available literature~\cite{dasilva2007formation, liu2013structural, panchal2011zircon, cococcioni2005linear}. To further improve the description of $d$ and $f$ behavior, the standard Heyd-Scuseria-Ernzerhof hybrid functional (HSE06) approach, i.e., mixing DFT with 20\% exact exchange is used. The phonon calculations are carried out within the density functional perturbation theory using \textbf{q}-mesh of $2\times 2 \times 2$ and zone-center lattice vibrations are analyzed to identify Raman modes~\cite{kroumova2003bilbao}. The exchange interactions between Ce atoms with ground state spin alignments are calculated using the Heisenberg spin Hamiltonian: $H_{\text{spin}}=-\sum_{i\neq j} J_1 \mathbf{S}_i \cdot \mathbf{S}_j - \sum_{k \neq l} J_2 \mathbf{S}_k \cdot \mathbf{S}_l$, where $J_1$ and $J_2$ represent the first- and second-nearest neighbor interactions. In CePO$_4$, there are two nearest neighbors of the same spin alignments and only one second nearest neighbor with opposite spin alignments, while in CeVO$_4$, there are four nearest neighbors with opposite spin alignments and eight second nearest neighbors with the same spin alignments (Fig.~\ref{fig1}).  Thus, the total energies of different spin configurations in CePO$_4$ are expressed as: ${E}_{FM} = {E}_0 - (2{J}_1 + {J}_2)S^2$, $E_{AFM1} = {E}_0 - (2{J}_1 - {J}_2){S}^2$, and $E_{AFM2} = {E}_0 - (-2{J}_1 - {J}_2){S}^2$ and the exchange parameters \({J}_1 \) and \({J}_2 \) are then derived as: ${J}_1 = ({E}_{AFM2} - {E}_{AFM1})/4S^2$ and ${J}_2 = (E_{\text{AFM1}} - E_{\text{FM}})/2S^2$. Similarly, in CeVO\(_4\), ${E}_{\text{FM}} = {E}_0 - (4{J}_1 + 8{J}_2)S^2$, ${E}_{\text{AFM1}} = {E}_0 - (-4{J}_1 + 8{J}_2)S^2$, and ${E}_{\text{AFM2}} = {E}_0 + 8{J}_2 {S}^2$, and $J_1 = (E_{\text{AFM1}} - E_{\text{FM}})/8S^2$ and  $J_2 = (2E_{\text{AFM2}} - E_{\text{AFM1}} - E_{\text{FM}})/16S^2$.

\section{Result and discussion}
\subsection{Crystal structure and magnetic ground state}
\begin{figure*}[!ht]
\centering	
\includegraphics[width=0.90\textwidth]{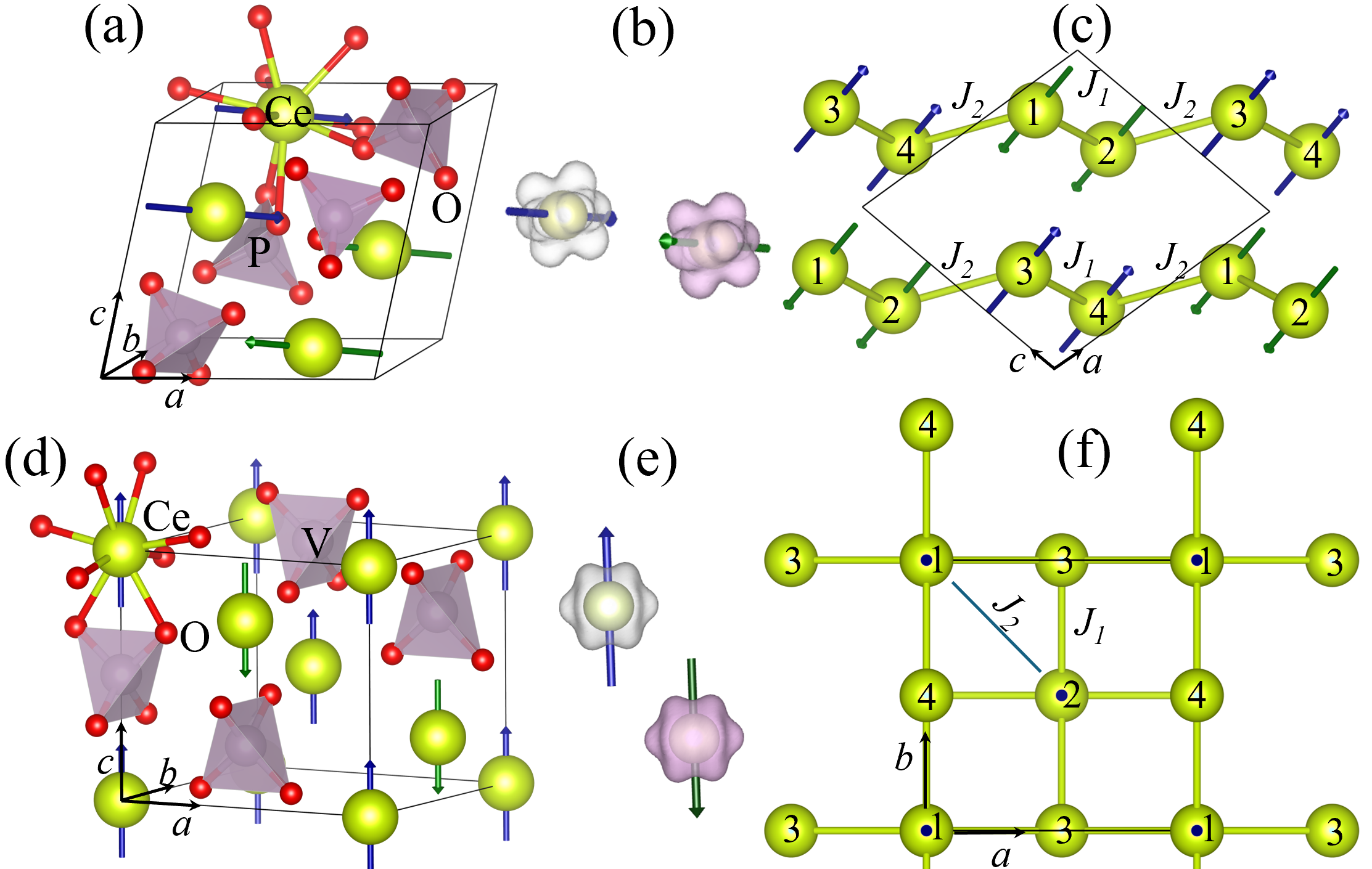}\hfill	
\caption{Crystal structures [(a) and (d)], electron spin density distributions of Ce in spin-up and spin-down directions [(b) and (e)], and exchange interactions [(c) and (f)] in the AFM monoclinic CePO$_4$ and AFM tetragonal CeVO$_4$. Blue and green arrows represent spin-up and spin-down Ce atoms in the AFM ground state structures. The P(V)O$_4$ tetrahedra are shown in both structures along with Ce atom coordinating to nine (eight) O atoms in CePO$_4$ (CeVO$_4$) [(a) and (d)]. The direction dependent total energy calculations, including spin orbit coupling, show that the displayed spin configurations, (110) in CePO$_4$ and (001) in CeVO$_4$, have lower total energy to form AFM ground state structures. The spin density contours [(b) and (e)] are drawn with an isosurface value of 0.01 e/\AA$^3$). $J_1$ and $J_2$ represent nearest neighbor and next nearest neighbor exchange interactions between Ce spins, respectively.}
\label{fig1}
\end{figure*}
CePO$_4$ and CeVO$_4$ belong to a family of wolframite-type oxides. The former crystallizes in two sub-structures: rhabdophane-type hydrated-hexagonal and monazite-type monoclinic ($P2_1/c$) and the latter crystallizes in a zircon-type tetragonal ($I4_1/amd$) structure at ambient conditions~\cite{savchyn2012vibrational, sisira2017microstructural}. In CePO$_4$, an irreversible phase transition between hydrated-hexagonal to a thermally stable monoclinic phase, induced either by doping or by adjusting the initial PO$_4^{3-}$/Ce$^{3+}$ ratio accompanied by dehydration at temperature above 400~$^{\circ}$C, is reported~\cite{allen2021tuning, fang2003systematic, sisira2017microstructural, lucas2004rare, sivaraman2024ceo2}. In both phases, every Ce atom is coordinated to nine oxygen atoms, and that of P atoms form PO$_4$ tetrahedra. The Ce atoms in the hexagonal phase are located on a six-sided double helix along the $c$ axis, while in the monoclinic structure, CeO$_9$ polyhedra are edge shared with PO$_4$ tetrahedra along the $c$-axis (Fig.~\ref{fig1}). We note that all nine different Ce-O bond lengths range from 2.474~\AA~to 2.803~\AA~and four different P-O bond lengths range from 1.546~\AA~to 1.564~\AA~in the monoclinic structure. The total energy calculations, with FM and AFM configurations in the monoclinic structure, predict AFM1 configuration (spin up for Ce1 and Ce2 and spin down for Ce3 and Ce4) with the lowest energy indicating it as a magnetic ground state. The magnetic ground state has 41.4~meV lower energy as compared to the FM configuration, while it has 1.5~meV lower energy as compared to other competing AFM2 structure (spin up for Ce1 and Ce3 and spin down for Ce2 and Ce4). The calculated lattice parameters of the ground-state structure (AFM1), $a$ =6.893~\AA, $b$ = 7.101~\AA, $c$ = 6.498~\AA, and $\beta$ = 103.95$^o$, are in good agreement with the literature~\cite{beall1981structure, bao2009low}.


In CeVO$_4$, the combined temperature and pressure effect transforms the zircon-type tetragonal structure to a monazite-type monoclinic structure~\cite{jain2013commentary, yoshimura1969new, panchal2011zircon, errandonea2011situ}, which is similar to that of monoclinic CePO$_4$ structure. It is noted that the monazite-type monoclinic structure is a distorted zircon-type tetragonal structure with higher compactness because of higher coordination of the Ce$^{3+}$ cation resulting in the formation of CeO$_9$ polyhedra instead of CeO$_8$ dodecahedra. Here, the tetragonal structure is composed of alternating edge-sharing CeO$_8$ dodecahedra and VO$_4$ tetrahedra forming chains parallel to the $c$~axis. The eight nearest neighbors of Ce are divided into two groups based on Ce-O bond lengths of 2.397~\AA~and 2.509~\AA. On the other hand, the V-O bonds in the VO$_4$ tetrahedra maintain the same length of 1.745~\AA. The standard density functional calculations neither produce magnetic ground state nor the experimentally measured band gap in CeVO$_4$~\cite{dasilva2007formation, liu2013structural, yoshimura1969new} due to V atom forming a correlated oxide along with the Ce atom. Our total energy calculations, including electron correlation effect using both GGA+$U$ and HSE06, with FM and AFM configurations confirm that the AFM1 configuration is the magnetic ground state, similar to that of the CePO$_4$. Here, the AFM1 is lower in energy than the FM configuration by 11~meV and 8~meV from GGA+$U$ and HSE06 calculations, respectively, which are in a reasonable agreement with the earlier reported GGA+$U$ results~\cite{da2007formation}. Interestingly, a local spin density approximation incorporating Hubbard $U$ (LSDA+$U$) reports 200~meV energy difference, which is quiet high~\cite{liu2013structural}. This suggests that the use of on-site electron correlation parameter on the top of DFT functionals is sensitive to the magnetic ground state configuration and the HSE06 does produce the most reliable ground state. Additionally, the HSE06 calculations also produce the experimentally measured band gap in CeVO$_4$ (discussed later). The calculated lattice parameters of the ground-state structure with AFM1 configuration, $a$ = 7.364 \AA~and $c$ = 6.484~\AA, are in good agreement with the literature~\cite{panchal2011zircon, liu2013structural}.



\subsection{Electronic structure, optical spectra, and magnetic interactions}

\begin{figure}[!ht]
\centering	
\includegraphics[width=0.49\textwidth]{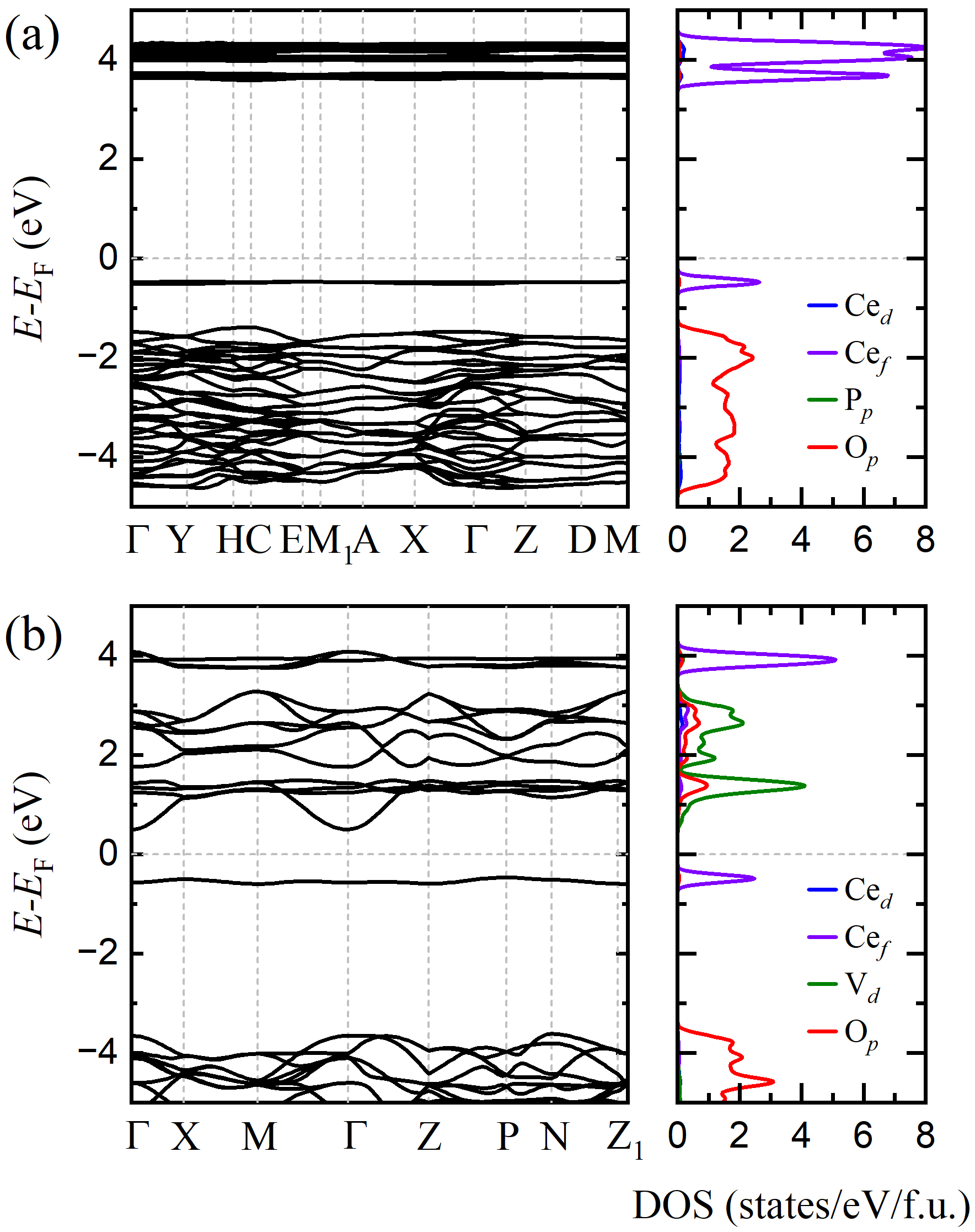}\hfill	
\caption{Electronic band structure and orbital projected DOS of (a) CePO$_4$  and (b) CeVO$_4$. The V $3d$ states, in CeVO$_4$, appear in both the occupied and unoccupied regime predominantly in the gap in between occupied and unoccupied Ce $4f$ states, while P $3p$ states of CePO$_4$ are located in high energy unoccupied regime.}
\label{fig2}
\end{figure}

The conventional local and semi-local exchange correlation functionals, mainly LDA and GGA in DFT significantly underestimate the band gap of strongly correlated oxides, primarily due to the delocalization or self-interaction error associated with the $4f$ electron states~\cite{perdew1981self, jones1989density}. The GGA calculations incorrectly identify both CePO$_4$ and CeVO$_4$ as metallic, whereas experimental evidence confirms them as wide band gap insulators~\cite{adelstein2011structure, garg2013phase}. Adding an onsite electron correlation, Hubbard $U$ in the DFT+$U$ approach, partially fixes this error and recovers the insulating state in both materials, as also suggested by previous reports~\cite{liu2013structural}. However, the value of $U$ and location and description of the 4$f$ states are not consistent~\cite{dasilva2007formation, liu2013structural}.  To overcome these issues, the calculations are performed with standard HSE06 functionals, in which a part of the exact exchange is admixed with the DFT, lead to a more accurate description of electronic structure~\cite{heyd2003hybrid}. HSE06 calculated band structures and DOS of the AFM monoclinic CePO$_4$ and tetragonal CeVO$_4$ (Fig.~\ref{fig2}) show band gaps that are close to the experimental values~\cite{liu2013structural, panchal2011zircon}. In CePO$_4$, the occupied and unoccupied Ce $4f$ states are localized and separated by 3.5~eV. The empty atomic-like levels in CePO$_4$ is consistent with what is anticipated for highly localized Ce $4f$ electrons.  While in CeVO$_4$, the occupied Ce $4f$ states are broader and the unoccupied Ce 4$f$ states spread over a large energy window hybridizing mainly with V 3$d$ states. This hybridization may affect the electronic transition between Ce 4$f$ states in the hybridization regime in CeVO$_4$ and the 4$f$ states behave as a weakly localized states. However, the center of gravity of unoccupied 4$f$ states in CeVO$_4$ lie above the V 3$d$ states in the conduction band because of the strong coulomb repulsion as also mentioned in Ref.~\cite{garg2013phase}.

\begin{figure}[!ht]
\centering	
\includegraphics[width=0.45\textwidth]{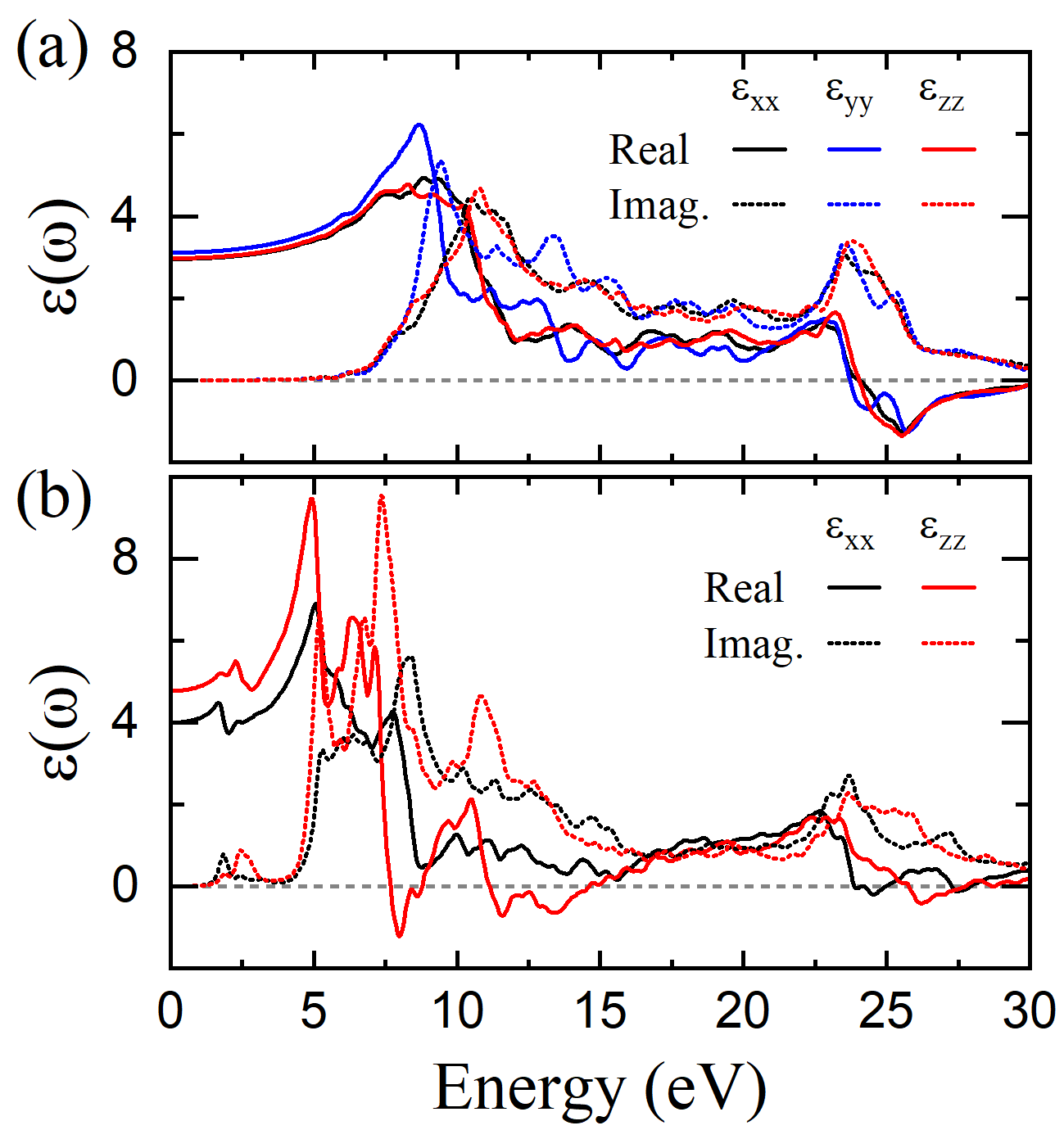}\hfill	
\caption{Real and imaginary dielectric constants as a function of photon energy for (a) CePO$_4$  and (b) CeVO$_4$.}
\label{fig3}
\end{figure}


The real and imaginary part of the polarization direction dependent complex dielectric constants of CeP(V)O$_4$ are calculated (Fig \ref{fig3}) and analyzed to understand the absorption behavior. The static dielectric constants close to 4 in all three directions in CePO$_4$ and 5.2 along the x/y and 6.2 along the z directions in CeVO$_4$ are identified, which characterize the degree of polarization under an electric field. These calculated values indicate moderate polarizability, good insulating properties, and a fairly high refractive indices, potentially fulfilling the requirement of photonic interface in both CeP(V)O$_4$. Further, the real part of the dielectric constant drops close to zero around 12~eV and crosses zero around 21~eV in CePO$_4$, while the y-component crosses zero around 6~eV, and x/y-components cross only around 21~eV in CeVO$_4$. These crossings are accompanied by peaks in the imaginary part, indicating a potential for plasmonic resonance, consistent with phenomena observed in epsilon-near-zero materials~\cite{liberal2017near}. The imaginary part is mainly characterized by a few significant absorption peaks with strong absorption due to electronic transitions and notable differences between polarization directions. For instance, the absorption peaks become prominent around 8.5~eV along the x/z-direction and 7.5~eV along the y-direction in CePO$_4$ and 6.5~eV along the x/y-direction and 3.5~eV along the z-direction in CeVO$_4$. These absorption peaks are associated with the inter-band (valence band to the conduction band) electronic transitions in the ultra-violet region. Here, the absorption peaks are indeed arising from transitions in the 4$f$ shell of Ce$^{3+}$ ions, as can also be seen form the electronic band structure (Fig.~\ref{fig2}).

The presence of mixed valance states (in between 3+ and 4+) of Ce in CePO$_4$ and CeVO$_4$ contribute to the interplay between localized and delocalized magnetism, exhibiting exotic magnetic and quantum behaviors. Due to the spin-orbit interaction, the Ce 4$f$ states contribute to both spin and orbital magnetic moments. Ideally, following Hund's rule, Ce in 3+ state should have orbital angular momentum, $L$ = 3 (orbital magnetic moment $\mu_L$ = 3 $\mu_B$), and spin angular momentum, $S$ = 1/2 (spin magnetic moment $\mu_S$ = 1 $\mu_B$), giving rise to total angular momentum, $J$ = $L - S$ = 5/2 (total magnetic moment $\mu_{total}$ = 2 $\mu_B$). In CePO$_4$, without SOC, the spin moment is 1~$\mu_B$, however, with the SOC along the z-direction, the point group symmetry $S_3$ changes to $C_1$. As a result, the Ce site splits into two non-equivalent sites with slightly different spin moments in the z-direction (0.7~$\mu_B$ and 0.8~$\mu_B$). Interestingly, small spin moment is also induced in other magnetization directions ($\mu_x \sim$0.5~$\mu_B$ and $\mu_y \sim$0.2~$ \mu_B$). The maximum orbital moment obtained for Ce 4$f$ in CePO$_4$ is 1 $\mu_B$ in the z-direction, which implies that the moment is indeed quenched by $\sim$2 $\mu_B$. We here note that, similar to the spin moments, the orbital moments are also induced  in other directions with values lower than 1 $\mu_B$. In CeVO$_4$, the spin moment is 1 $\mu_B$ without SOC. While applying SOC along the z-direction, the point group symmetry remains unchanged producing the same 1 $\mu_B$ spin magnetic moment. Interestingly, the orbital moment comes out to be 1.8 $\mu_B$ along the z-direction, which indeed shows lower level of orbital moment quenching in CeVO$_4$ as compared to CePO$_4$.

The magnetism of CeP(V)O$_4$ is primarily governed by the Ce 4$f$ electrons with highly anisotropic distribution of the spin density [Fig. \ref{fig1} (b) and (e)]. As a result, direction-dependent magnetic properties arise from the varying exchange interaction energies between neighboring Ce atoms. For example, calculated nearest neighbor ($J_1$) and next nearest neighbor ($J_2$) exchange interaction energies between Ce spins, respectively, are~0.89~meV and~-27.38~meV in CePO$_4$ and -11.1~meV and -3.8~meV in CeVO$_4$. These exchange interaction energies support the AFM1 ground states in both CePO$_4$ and CeVO$_4$, also identified from the total energy calculations. In addition, direction-dependent total energy calculations, incorporating SOC further confirm the variations in energetics as a function of crystallographic orientation. With the spin-quantization axis fixed to (001) direction and varying the on-site magnetic moment directions to (001), (100), (110), and (111), respectively, the magnetic moment aligned along the (110) direction is found lower in energy by 2~meV with the (111), 180~meV with the (001), and 201~meV with the (100) directions in CePO$_4$. In CeVO$_4$, the magnetic moment aligned along the (001) direction is lower in energy by 18~meV with the (100), 32~meV with the (110), and 250~meV with the (111) directions. Therefore, the easy magnetization direction (Fig.~\ref{fig1}) and hence the spin projection along the quantization axis are different for CePO$_4$ and  CeVO$_4$. This analysis accounts for the interplay between the anisotropic spin density and the spin-orbit interaction, providing detailed insights into the direction dependent magnetic behavior. 

\subsection{Phonon dispersion and Raman-active modes}
\begin{figure}[!ht]
\centering	
\includegraphics[width=0.45\textwidth]{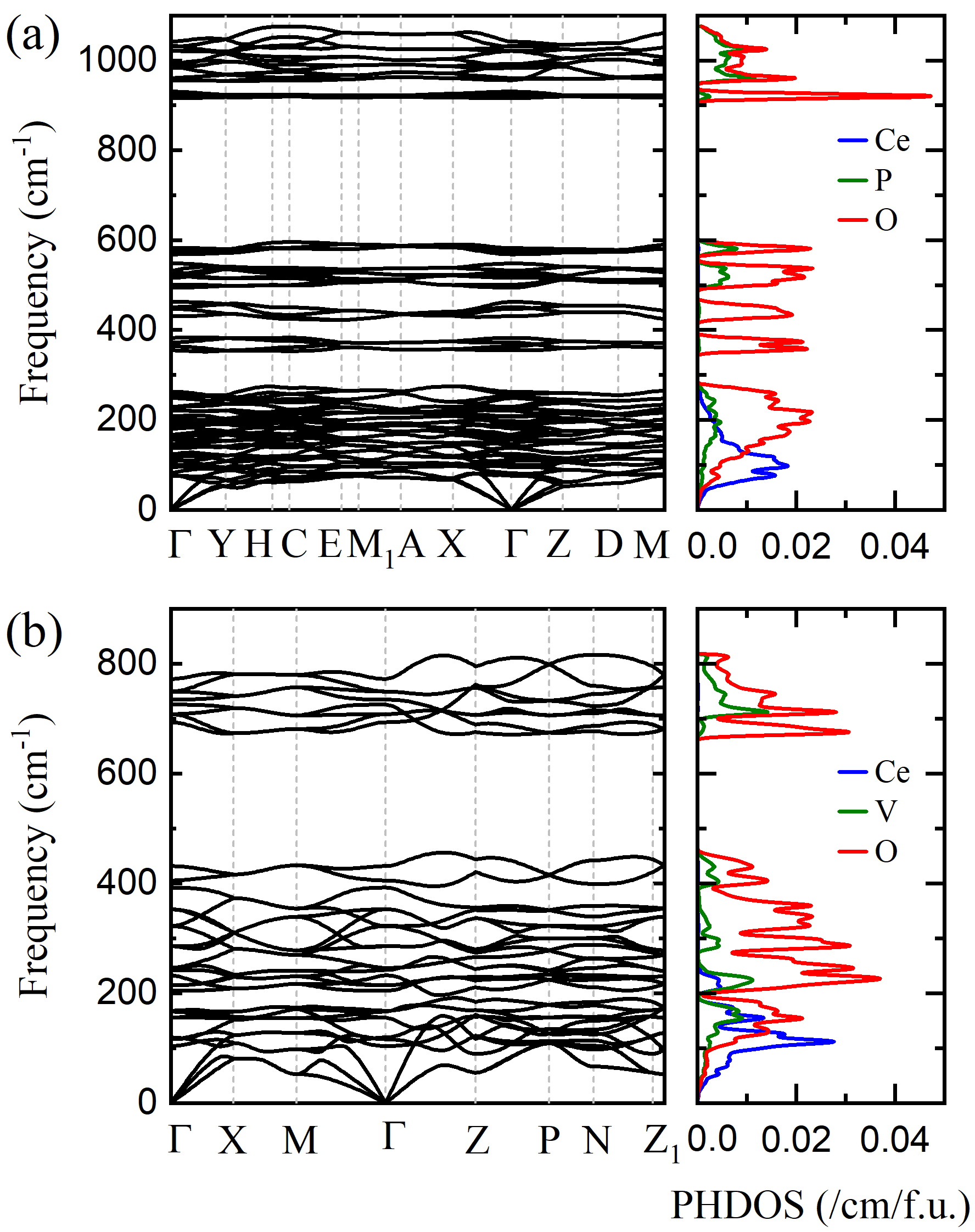}\hfill	
\caption{Phonon dispersion and atom projected phonon DOS of (a) CePO$_4$  and (b) CeVO$_4$.}
\label{fig4}
\end{figure}
To understand the correlation between phonon dynamics and Raman modes, as well as their connection to quantum information processing in CePO$_4$ and CeVO$_4$, we calculate and analyze their phonon frequencies, phonon DOS (PHDOS) and Raman-active modes (Fig.~\ref{fig4} and Table~\ref{table1}). The acoustic phonon modes in both CePO$_4$ and CeVO$_4$ show conventional characteristics, however, the optical phonon modes in CePO$_4$ exhibit flatter dispersion compared to that of CeVO$_4$ across the entire vibrational spectrum. Here, the complete phonon spectrum can be divided into three parts: low frequency regime (below 300 cm$^{-1}$ in CePO$_4$ and below 200~cm$^{-1}$ in CeVO$_4$), intermediate frequency regime (320-600~cm$^{-1}$ in CePO$_4$ and 200-450~cm$^{-1}$ in CeVO$_4$), and high frequency regime (above 930~cm$^{-1}$ in CePO$_4$ and above 670~cm$^{-1}$ in CeVO$_4$). Two clear phonon gaps of $\sim$40~cm$^{-1}$ and $\sim$320~cm$^{-1}$ are observed in  between these regions in CePO$_4$, while in CeVO$_4$, the lower gap is almost closed and the upper gap is only $\sim$200~cm$^{-1}$. The low frequency modes comprise a mixture of vibrations involving all three types of atoms predominantly by Ce atoms, while the intermediate and high frequency modes are associated mainly with the vibrations of P and O atoms in CePO$_4$ and V and O atoms in CeVO$_4$. Importantly, the vibrations of O atoms are extended across the entire phonon spectrum constituting an essential component of the CeO$_{9}$ and PO$_4$ polyhedra in CePO$_4$ and CeO$_8$ and VO$_4$ polyhedra in CeVO$_4$. This leads to collective oscillations with mixed vibrational modes due to the involvement of polyhedra as also reported in Cu doped lead apatite complex structure~\cite{paudyal2024implications}. The softening of acoustic modes at the $M$ and $Z$ points of the Brillouin zone in CeVO$_4$ indicate a possibility of soft phonon mode driven structural phase transition~\cite{ying2018unusual, kafle2020electronic}. Indeed, with applied pressure and/or temperature the tetragonal phase transforms to the monoclinic phase as reported in the literature~\cite{panchal2011zircon, errandonea2011situ}.

The phonon dispersion reveals the presence of both the Raman- and infrared-active modes in CePO$_4$ and CeVO$_4$ based on the crystal symmetry. The point group of CePO$_4$ is $C_{2h}$, which possesses a two-fold rotation axis ($C_2$) along with reflection planes ($\sigma_h$) perpendicular to the rotation axis. On the other hand, the point group of CeVO$_4$ is $D_{4h}$ which includes all the symmetry elements of the $C_{2h}$ along with a four-fold rotation axis $C_4$ and diagonal reflection planes ($\sigma_d$). Additionally, $D_{4h}$ contains a center of inversion and possesses inversion symmetry. For the monoclinic $C_{2h}$ structure of CePO$_4$, 36 Raman-active phonon modes are expected with 18$A_g$ and 18$B_g$ irreducible representations, while in the tetragonal $D_{4h}$ structure of CeVO$_4$, 12 Raman-active phonon modes are expected with 2$A_{1g}$, 4$B_{1g}$, $B_{2g}$, and 5$E_g$ irreducible representations. We note that the doubly degenerate E$_g$ modes in the tetragonal phase split to form  $A_g$/$B_g$ modes in the monoclinic phase due to symmetry reduction. Here, the P(V)O$_4$ tetrahedron has four distinct internal and two external vibrational modes that involve both Ce atoms and P(V)O$_4$ polyhedra as also identified in the literature for similar material systems~\cite{guedes2001raman, santos2007raman, lalla2021structural, geisler2016evidence}. The internal modes can be either symmetric/anti-symmetric stretch ($\nu_1$/$\nu_3$) or symmetric/anti-symmetric bending ($\nu_2$/$\nu_4$) and the external modes are translation (T)- or rotation (R)-like vibrations. The calculated Raman active phonon frequencies are in good agreement with the previous neutron inelastic and Raman-scattering measurements~\cite{panchal2011zircon, silva2006vibrational} (Table~\ref{table1}). The low-frequency Raman modes involve both Ce and P(V)O$_4$ tetrahedra and can be characterized by external T- and R-like vibrations. In the tetragonal CeVO$_4$, the phonon mode characteristics (Table~\ref{table1}) match well with previous studies, while in the monoclinic CePo$_4$, these modes are complex due to the presence of mixed vibrations. However, we are able to characterize them with the help of their eigenvectors. Here the nine internal modes for each $A_g$ and $B_g$ symmetry above 300~cm$^{-1}$ involve the PO$_4$ tetrahedra, while 18 modes below 300~cm$^{-1}$ involve external T- and R-like modes. Interestingly, the high-frequency vibrations of O atoms corresponding to the anti-symmetric stretch ($\nu_3$) in CePO$_4$ are found highly anisotropic as compared to that in CeVO$_4$, indicating a stronger PO$_4$ bonding than that of VO$_4$ tetrahedra. This causes difference in crystal-field effects in CePO$_4$ and CeVO$_4$ (discussed later).

\vspace{-0.3cm}
\begin{table}[ht!]
\centering
\caption{Calculated and experimental Raman frequency mode frequencies. The symmetry assignment of the Raman modes are consistent with available literature~\cite{panchal2011zircon, guedes2001raman}, where $\mu_{1...4}$ correspond to the internal modes of P(V)O$_4$ tetrahedron and T (R) corresponds to the translational (rotational) motion involving both Ce and P(V)O$_4$ units.}
\begin{tabular}{l c c c | c c c c c}
\toprule
CePO$_4$ & Calc. & Calc. & Expt. & CePO$_4$ & Calc. & Calc. & Expt. \\
& FM &  AFM & \cite{silva2006vibrational}  &  & FM & AFM & \cite{silva2006vibrational}  \\
\hline
A$_g$ (T) & 88.1  & 95.2  & - & A$_g$ ($\nu_4$) & 590.0 & 591.0 & 588 \\
B$_g$ (T) & 94.7  & 103.9 & 88 & B$_g$ ($\nu_4$) & 591.3 & 591.9 & 618 \\
A$_g$ (T) & 101.8 & 104.9 & 100 & B$_g$ ($\nu_3$) & 909.5 & 913.1 & - \\
A$_g$ (T) & 132.8 & 127.1 & - & A$_g$ ($\nu_3$) & 915.9 & 919.7 & - \\
B$_g$ (T) & 142.3 & 139.0 & - & A$_g$ ($\nu_3$) & 945.0 & 947.0 & - \\
B$_g$ (T) & 144.6 & 144.8 & - & A$_g$ ($\nu_1$) & 967.1 & 966.6 & 970 \\
B$_g$ (R) & 147.7 & 149.8 & - & B$_g$ ($\nu_3$) & 976.1 & 976.7 & 990 \\
A$_g$ (R) & 153.6 & 153.0 & 152 & B$_g$ ($\nu_1$) & 1004.0 & 1004.6 & 1025 \\
B$_g$ (R) & 160.2 & 158.3 & 158 & A$_g$ ($\nu_3$) & 1010.0 & 1010.8 & 1054 \\
A$_g$ (T) & 170.0 & 167.7 & 172 & B$_g$ ($\nu_3$) & 1023.2 & 1022.2 & 1071  \\
A$_g$ (T) & 178.4 & 176.6 & 183 &  &  &  &  \\ \cline{5-8}
B$_g$ (T) & 203.9 & 208.4 & - & CeVO$_4$ & Calc. & Calc. & Expt.\\
A$_g$ (T) & 205.1 & 215.6 & 219  &          & FM   & AFM & \cite{panchal2011zircon} \\ \cline{5-8}
B$_g$ (T) & 218.3 & 228.3 & 227  & E$_g$ (T)    & 110.8 & 117.6 &  - \\	
B$_g$ (T) & 239.3 & 245.2 & 254  & B$_{1g}$ (T) & 131.6 & 131.1 &  124.4 \\
A$_g$ (R) & 240.3 & 249.5 & -  & E$_g$ (T)    & 166.4 & 171.6 &  - \\
B$_g$ (R) & 257.9 & 269.3 & -  & E$_g$  (R)   & 263.2 & 266.7 &  234.1\\
A$_g$ (R) & 264.3 & 272.4 & 270  & B$_{1g}$ (T) & 247.4 & 248.7 &  - \\
B$_g$ ($\nu_4$) & 368.6 & 368.1 & -  & B$_{2g}$ ($\nu_2$) & 249.4  & 248.1 &  261.9 \\
A$_g$ ($\nu_4$) & 387.7 & 386.5 & 396  & A$_{1g}$ ($\nu_2$) & 321.1  & 322.3 &  - \\
A$_g$ ($\nu_2$) & 442.2 & 443.4 & 414  & E$_g$ ($\nu_4$)    & 372.6  & 373.4 &  381.1 \\
B$_g$ ($\nu_2$) & 486.1 & 488.7 & 467  & B$_{1g}$ ($\nu_4$) & 429.2  & 430.8 & 468.9 \\
A$_g$ ($\nu_2$) & 506.9 & 508.8 & -  & B$_{1g}$ ($\nu_3$) & 779.0  & 781.9 & 789.1 \\
B$_g$ ($\nu_2$) & 529.7 & 530.6 & 535  & E$_g$ ($\nu_3$)    & 784.1  & 784.5 &  801.3 \\
A$_g$ ($\nu_4$) & 541.4 & 543.0 & -  & A$_{1g}$ ($\nu_1$) & 836.3  & 838.6 & 864.3 \\
B$_g$ ($\nu_4$) & 558.5 & 557.8 & 568  &  &  &  &  \\
\hline
\hline
\end{tabular}
\label{table1}	
\end{table}
\vspace{0.35cm}


\subsection{Spin-phonon coupling}
We now analyze spin–phonon coupling strength associated with Raman active modes. In the magnetically ordered structure, the phonon frequency shift due to the spin-lattice coupling quantifies spin-phonon coupling strength. As expected, there exists an obvious dependency between the phonon spectrum and spin arrangement exhibiting a strong correlation between lattice vibration and electron spin. The shift in the phonon frequency due to spin-phonon interaction can be described phenomenologically by employing a static spin-spin correlation function, $\lambda \langle S_i.S_j \rangle $, $\lambda$ being the spin-phonon coupling constant. The optical Raman active modes at the $\Gamma$ point of the Brillouin zone (Table~\ref{table1}) show a noticeable differences, especially in the low frequency regime with respect to the spin ordering in both CePO$_4$ and CeVO$_4$. For instance, in CePO$_4$, an $A_g$ mode located at 205.1 cm$^{-1}$ in the FM and 215.6 cm$^{-1}$ in the AFM spin states and a $B_g$ mode located at 257.9 cm$^{-1}$ in the FM and 269.3 cm$^{-1}$ in the AFM spin states have significant differences. Here, the $A_g$ mode corresponds to the internal translational motion, while the $B_g$ mode involves the rotational movement of the PO$_4$ tetrahedra. Similarly, in the CeVO$_4$, an $E_g$ mode located at 110.8 cm$^{-1}$ in the FM and 117.6 cm$^{-1}$ in the AFM spin states exhibit larger differences. This $E_g$ mode corresponds to the internal translational motion involving both Ce and VO$_4$ tetrahedra. For both CePO$_4$ and CeVO$_4$ with spin half 4$f$ (3+ oxidation state) for Ce, the spin-spin correlation function is 1/4. The largest phonon frequency change results in an spin-phonon coupling constant $\lambda$ of 45.6 cm$^{-1}$ in CePO$_4$ and 27.2 cm$^{-1}$ in CeVO$_4$, which are relatively larger than that observed in 5$d$ oxides~\cite{calder2015enhanced, zhang2021first, kim2020spin}, implying that the phonon dynamics is strongly influenced by the spin ordering. Such a novel collective excitation involving the lattice, spin, and charge degrees of freedom and the ability to control phonons utilizing these parameters makes a potential foundation not only for spintronic and optoelectronic science but also for quantum information science.

\subsection{Crystal field excitation and quantum implications for quantum transduction and memory}
The Ce$^{3+}$ ion in CeP(V)O$_4$ has only one 4$f$ electron and the spin-orbit splitting produces $^2F_{5/2}$ ground state and $^2F_{7/2}$ excited state. Both ground and excited states should be Kramers degenerate, since $J$ has fractional values. Three Ce$^{3+}$ ions occupy $C_1$ sites in CePO$_4$, while two Ce$^{3+}$ ions occupy $D_{2d}$ sites and are connected by an inversion center in CeVO$_4$. To investigate the crystal field split 4$f$ states of Ce in CePO$_4$ and CeVO$_4$, we solve an effective Hamiltonian using electronic structure calculated crystal field coefficients (CFCs)~\cite{limbu2025ab,limbu2025abc}. Here, for CFCs calculations, we keep the 4$f$ electrons of Ce in the core and perform non-magnetic calculations to generate the self consistent charge densities and local potentials~\cite{novak2013crystal}. 
In both the $C_1$ site symmetry in CePO$_4$ and the $D_{2d}$ site symmetry in CeVO$_4$, three doubly degenerate $Z_1$, $Z_2$, and $Z_3$ ground state multiplets and four doubly degenerate first excited state $Y_1$, $Y_2$, $Y_3$, and $Y_4$ multiplets are identified (Table~\ref{table2}). The irreducible representation of these ground and first excited state multiplets of CePO$_4$ is $\Gamma_2$ (Table~\ref{table2}) as also suggested by Refs.~\cite{goodman1991crystal,bradley1976p}. However, the irreducible representation of the ground and first excited state multiplets of CeVO$_4$ are $\Gamma_6$ and $\Gamma_7$ (Table~\ref{table2}) as also suggested by Refs.~\cite{bhattacharyya2022crystal,bradley1976p}. The calculated electronic transition from $Y_1$ to $Z_1$ in CePO$_4$ and CeVO$_4$ are 2266~cm$^{-1}$ and 2228~cm$^{-1}$, respectively. The $Y_1$ to $Z_1$ transition in CeVO$_4$ closely agrees with the available literature value in the $D_{2d}$ site symmetry of Ce$^{3+}$ in NaCeO$_2$~\cite{bhattacharyya2022crystal}.
\begin{table}[ht!]
\centering
\caption{Calculated energy level structure of Ce$^{3+}$ in CePO$_4$ and CeVO$_4$. The $\Gamma$s are irreduciable representations.}
\begin{tabular}{c c l@{} c}
\hline
\hline
\multirow{2}{*}{$^{2S+1}F_J$}~ & \multirow{2}{*}{Multiplets} & ~~~CePO$_4$~~~ & ~~~CeVO$_4$~~~ \\
  & & ~~~(cm$^{-1}$)~~ & (cm$^{-1}$) \\
\hline
\multirow{3}{*}{$^2F_{5/2}$} & Z$_1$ & \multirow{3}{*}{$\left.\begin{array}{l}
                ~~~~0 \\
                ~~~210 \\
                ~~~581 
                \end{array}\right\rbrace \Gamma_2 $} & ~~~~~0 ~~~~$\Gamma_6$ \\
             & Z$_2$ &  & \multirow{2}{*}{$\left.\begin{array}{l}
                ~~~16 \\
                ~~192 \end{array}\right\rbrace \Gamma_7 $}  \\
             & Z$_3$ & \\\\
\multirow{4}{*}{$^2F_{7/2}$} & $Y_1$ & \multirow{4}{*}{$\left.\begin{array}{l}
                ~~2266 \\
                ~~2392 \\
                ~~2596 \\
                ~~2944
                \end{array}\right\rbrace \Gamma_2 $} &  \multirow{2}{*}{$\left.\begin{array}{l}
                2228 \\
                2306 \end{array}\right\rbrace \Gamma_6 $} \\
             & Y$_2$ &  \\
             & Y$_3$ &  & \multirow{2}{*}{$\left.\begin{array}{l}
                2327 \\
                2507 \end{array}\right\rbrace \Gamma_7 $}  \\
             & Y$_4$ & \\\\
\hline
\hline
\end{tabular}
\label{table2}	
\end{table}

The crystal symmetry and the site-symmetry of Ce atoms in the CeP(V)O$_4$ exhibit distinct quantum behaviors and utilities for spin-photon interfaces. Of the two different site symmetries of Ce in monoclinic CePO$_4$ and tetragonal CeVO$_4$, the $C_1$ site symmetry of Ce in CePO$_4$ lacks inversion and is relevant for quantum memories and $D_{2d}$ including inversion symmetry of Ce in CeVO$_4$ is relevant for quantum transduction.  The $C_1$ site, lacking inversion symmetry, has a static electric dipole moment, making CePO$_4$ advantageous for quantum memory where Ce-ion optical transitions can be tuned in and out of resonance with an optical cavity using an electric field. We note that this symmetry breaking modifies selection rules for optical transitions, enhancing spin-photon interactions and enabling efficient optical control of quantum states~\cite{goldner2015book, bhandari2023distinguishing}. However, the $D_{2d}$ site, without a static electric dipole moment in CeVO$_4$, is better suited for narrow-linewidth applications in quantum transduction. Inversion symmetry in quantum systems aids quantum transduction by enhancing coherence, controlling selection rules for transitions, optimizing phase matching conditions in nonlinear optics, facilitating controlled spin-photon interactions, and maintaining well-defined energy levels through symmetric crystal fields~\cite{anderson2023embracing}. These factors contribute to the efficient and coherent transfer of quantum information between different physical modalities, which is essential for building robust and high-fidelity quantum transduction devices.

\section{Conclusion}
$Ab~initio$ calculations identify AFM ground states in structurally preferred monoclinic CePO$_4$ and tetragonal CeVO$_4$ exhibiting strongly localized occupied and unoccupied Ce $4f$ states with Ce $4f$ - Ce $4f$ transition characteristics. In CePO$_4$, O $2p$ and P $3p$ states hybridize negligibly with Ce $4f$ contrasting with the CeVO$_4$ in which $3d$ states of V and $2p$ states of O appear as extended states in between the occupied and unoccupied Ce $4f$ states. The hybridization between these extended states and their location in between the prominent localized occupied and unoccupied Ce $4f$ DOS peaks in CeVO$_4$ may become a barrier to electronic transition fidelity as compared to that of CePO$_4$. Here, phonon calculations and analysis identify and differentiate Raman active phonon modes along with spin phonon coupling of cerium in both CePO$_4$ and CeVO$_4$ that ultimately lead to different $4f$ ground state crystal field multiplets critical to accurate electronic transitions for foundational quantum transduction and memories. The $C_1$ site symmetry of Ce in CePO$_4$, which lacks inversion symmetry, is relevant for quantum memories and $D_{2d}$ Ce site with inversion symmetry in CeVO$_4$ is relevant for quantum transduction. 

\section*{Acknowledgements}
This work is primarily supported by the U.S. Department of Energy, Office of Science, Office of Basic Energy Sciences under Award Number DE-SC0023393. The spin phonon coupling  part of the work is supported as part of the Center for Energy Efficient Magnonics, an Energy Frontier Research Center funded by the U.S. Department of Energy, Office of Science, Basic Energy Sciences, under Award number DE-AC02-76SF00515. We acknowledge use of the computational facilities on the Frontera supercomputer at the Texas Advanced Computing Center via the pathway allocation, DMR23051, and the Argon high-performance computing system at the University of Iowa.

\bibliography{references}
\end{document}